\documentclass[pra, aps, twocolumn]{revtex4}

\usepackage{graphicx, epsfig}
\usepackage{amssymb,amsmath}

\def \hf{\tfrac{1}{2}}

\newcommand{\ra}{\rightarrow}   

\def\lba{\left(}    \def\rba{\right)}
\def\lbc{\left[}    \def\rbc{\right]}

\renewcommand{\vr}{{\bf r}}

\begin{document}

\title{A vortex dipole in a trapped two-dimensional Bose-Einstein condensate}

\author{Weibin Li}

\author{Masudul Haque}

\author{Stavros Komineas}

\affiliation{Max Planck Institute for Physics of Complex Systems, N\"othnitzer
  Strasse 38, 
Dresden, Germany}

\date{\today}

%
%

\begin{abstract}

We study the conservative dynamics and stationary configurations of a
vortex-antivortex pair in a harmonically trapped two-dimensional Bose-Einstein
condensate.
We establish the conceptual framework for understanding the stationary states
and the topological defect trajectories, through considerations of different
mechanisms of vortex motion and the bifurcation of soliton-like stationary
solutions.  Our insights are based on Lagrangian-based variational
calculations, numerical solutions of both the time-dependent and
time-independent Gross-Pitaevskii equations, and exact solutions for the
non-interacting case.

\end{abstract}


\maketitle

\section{Introduction} 
Vortex-antivortex pairs or \emph{vortex dipoles} are ubiquitous in
two-dimensional (2D) Bose condensates.  They are the most natural excitations
when a 2D fluid flows past a barrier or through a disordered medium.  Such
creation of vortices and antivortices has been seen explicitly in recent
experiments with trapped atomic condensates
\cite{anderson-arizona_vortex-by-merging_PRL07}.  In 2D, vortex dipoles can be
thermally created.  This is important for the thermodynamics of a 2D
superfluid, \emph{e.g.}, for the Kosterlitz-Thouless phase transition recently
studied experimentally \cite{Dalibard_KT-expts}.
Given their pervasiveness and the current experimental interest with
harmonically trapped condensates, a thorough study of vortex dipoles in trapped
condensates is clearly important.

In this Article, we use a combination of methods to analyze vortex-antivortex
states and dynamics in a circularly trapped 2D condensate, as described by the
time-dependent Gross-Pitaevskii equation (GPE).
First, we use a variational approach to the GPE that can be solved almost
analytically, providing qualitative insight.
%
%
These insights are explored and complemented by explicit numerical evolutions
of the time-dependent GPE, exact solutions for the non-interacting case, and a
numerical procedure for accessing stationary solutions of the time-independent
GPE.
%
%
Our results indicate that the dynamics of a vortex-antivortex pair in a trap
is unexpectedly rich.  In this Article we focus on a subset of possible
dynamics, namely, the pair trajectory for cases where the pair is initially
placed symmetrically on opposite sides of the trap center.  
%

The physics behind our characteristic pair trajectories involves two effects.
First, a vortex dipole in a large fluid body is a \emph{propagating} object,
\emph{i.e.}, the vortex and antivortex propel each other in a direction
perpendicular to the line joining them.  This has been studied in condensates
\cite{Fetter_PRA65-I, JonesRoberts_JPhysA82} and is also common experience for
ordinary fluids.
In a trapped non-rotating condensate, there is an additional effect.  An
off-center vortex is known to precess around the center of the trap
\cite{Rokhsar_PRL97, Fetter_PRL84}; we can regard this as inhomogeneity-driven
motion.  For a vortex dipole in a trap, each of the pair is driven by the
inhomogeneity, in a direction opposite to the mutually-driven
motion.
One of the two types of motion, mutually driven or inhomogeneity-driven, can
dominate, depending on the vortex-antivortex distance.  This leads to the
remarkable fact that a vortex dipole with the same sense (same dipole
direction) can be propagating in one of two opposite directions, depending on
the dipole separation.  A similar situation holds for vortex rings in 3D
trapped condensates \cite{KomineasPapanicolaou_LaserPhys04,
KomineasBrand_PRL05}.
Another result of the competition described above is that the two effects can
in some cases exactly balance each other, leading to \emph{stationary
configurations} of vortex pairs.  In this Article, we will examine in detail
effects of this competition, for a vortex dipole in a circularly trapped
condensate.

The conservative dynamics presented here should be accessible experimentally,
\emph{e.g.}, by creating a vortex-antivortex pair by phase imprinting,
analogous to the first creation of a vortex in a laser-cooled condensate
\cite{Cornell-expt_PRL99}. In contrast, the recent experiments
\cite{anderson-arizona_vortex-by-merging_PRL07, Dalibard_KT-expts} deal with
\emph{dissipative} physics involving creation and destruction of vortex
dipoles.
While we have omitted dissipative considerations in this study, we will see
that the conservative dynamics is extremely rich by itself.

Theoretically, vortex dipoles in trapped condensates have appeared as a
peripheral topic in numerical studies of 3D vortex rings
\cite{JacksonAdams_PRA99, KomineasPapanicolaou_LaserPhys04,
RuostekoskiDutton_PRA05, Komineas_review07, KomineasBrand_PRL05}, which may be
regarded as 3D analogs of 2D vortex dipoles.  Refs.~\cite{crasovan-etal_PRA03,
Helsinki_stationary-clusters_PRA05} have considered stationary vortex dipole
states, which we will further examine and interpret,
Ref.~\cite{ZhouZhai_PRA04} has studied energetics, while
Ref.~\cite{Helsinki_stationary-clusters_PRA06} presented some density
dynamics.  Very recently, Ref.~\cite{KleinJakschZhangBao_PRA07} has studied
some vortex dipole dynamics in the weak-interaction region.

Sec.~\ref{sec_variational} presents results from our variational calculation.
Although the variational results cannot be expected to be quantitatively
accurate, the calculation provides physical intuition.  In particular, the
most prominent feature of the vortex dipole trajectories, involving each
singularity revolving around a stationary point, already emerges from this
calculation.  In Sec.~\ref{sec_noninteracting} we present relevant exact
results for the non-interacting case, focusing again on defect trajectories.
Sec.~\ref{sec_soliton} presents numerical solutions of the time-independent
GPE.  This provides valuable insight into the differences observed in the
dynamics for small interactions and large interactions.  In
Sec.~\ref{sec_tdgpe} we present direct numerical simulations of the
time-dependent GPE.  The earlier calculations (Secs.~\ref{sec_variational},
\ref{sec_noninteracting}, \ref{sec_soliton}) allow us to interpret the results
of the time-dependent simulations; Secs.~\ref{sec_densities} and
\ref{sec_discussion} present the interpretations and point out open questions.

\emph{Gross-Pitaevskii Equation.} \ 
We study a low-temperature Bose-Einstein condensate in a harmonic trap.  From
the outset we start with a two-dimensional system confined to the $xy$ plane,
assuming that the dynamics in the $z$ direction is frozen out by tight
confinement which we do not treat explicitly.  We will consider circular
traps, and measure lengths in units of trap oscillator length and time in
units of inverse trapping frequency.  The condensate dynamics is given by the
time-dependent Gross-Pitaevskii equation \cite{pitaevskii-jetp13,gross-nc20}:
\begin{equation}
\label{gpe}
i\frac{\partial\psi(t)}{\partial t}
= -\hf\bigtriangledown^2\psi+V_{\rm tr}({\bf r}) \psi+g|\psi|^2\psi
\; .
\end{equation}
The trap potential is $V_{\rm tr}=\hf(x^2+y^2)$.
The effective 2D interaction strength $g$ characterizes the two-body
interaction, but also depends proportionally on the total particle number $N$
and inversely on the oscillator length.
%

\begin{figure}
\centering
\includegraphics*[width=0.95\columnwidth]{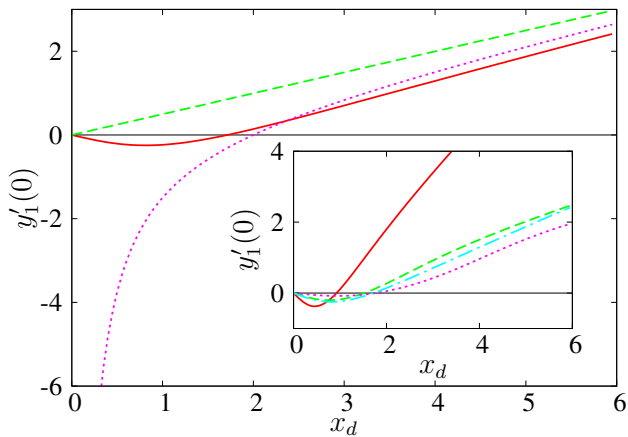}
\caption{  \label{fig_init-velocity-vs-xd}
(Color online.) Initial pair velocities versus dipole separation.
Thick-solid: Gaussian ansatz, $g=0.1$.  Dashed: a single vortex, same ansatz,
same $g=0.1$.  Dotted line: non-interacting condensate, $\hf{x_d}-2/x_d$. \;
Inset: Initial velocities versus $x_{\rm d}$, Solid, dashed and dotted-dashed
lines are from Thomas-Fermi ansatz,with $g=$ 1, 10, and 100.  Dotted line
is from gaussian ansatz, $g=1$.  All quantities are plotted in trap units.
}
\end{figure}

\section{Insights from variational calculation}   \label{sec_variational}

In this Section we will draw physical insights from a time-dependent
variational calculation.  This is not expected to give quantitative
predictions, but will provide valuable intuition about the stationary
solutions and characteristic trajectories of the vortex dipole.
We use the following class of simple variational wavefunctions:
\begin{equation}
\label{trialwf_general}
\psi ~=~ A(t)\; \lbc{z-z_1(t)}\rbc\lbc{z^*-z^*_2(t)}\rbc \; f_{\rm c}(|z|^2)
\; ,
\end{equation}
where $z=x+iy$ is the complex 2D position coordinate, $z_1$ and $z_2$ are
vortex and antivortex positions, and $A(t)$ is a normalization constant.  Both
Gaussian and Thomas-Fermi forms are used for the condensate shape function
$f_{\rm c}$.  Details of the formulation are in the Appendix.

Our form \eqref{trialwf_general} is amenable to exact treatment.  Its
disadvantage is that the vortices are `too big', being determined by the trap
size rather than the healing length $\xi$, which is reasonable only for small
interactions.  We note however that in two dimensions $\xi\sim{g}^{-1/4}$
decreases rather slowly with the interaction parameter, so that vortices
remain big even for relatively large values of $g$.  For the physics of
stationary states, our wavefunctions therefore give good answers up to large
$g$ (up to $g\sim{90}$).  The situation with dynamics is more complicated.

\begin{figure}
\centering
\includegraphics*[width=0.9\columnwidth]{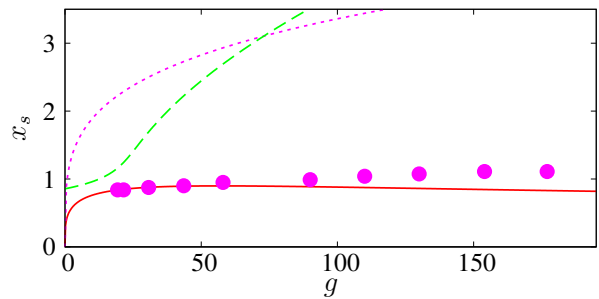}
\caption{  \label{fig_xs-vs-g}
(Color online.)  Half-distance between pair for stationary vortex dipole
configuration, $x_{\rm s}$, as function of $g$.  Solid line: from Thomas-Fermi
ansatz; dashed line from Gaussian ansatz; filled circles from time-independent
GPE.  The dotted line shows the Thomas-Fermi boundary of condensate.  
Lengths are plotted in trap units.
}
\end{figure}

\subsection{Stationary solution}  \label{sec_variational-stationary}

We restrict ourselves to initial positions of the vortex and antivortex which
are symmetric around the trap center; without loss of generality we place both
on the $x$-axis, $y_1(0)=y_2(0)=0$.  If we start with $x_2(0)=-x_1(0)<0$,
mutually-driven (inhomogeneity-driven) motion would tend to drive both defects
in the negative (positive) $y$ direction.  Mutually-driven
(inhomogeneity-driven) motion wins when the vortex and antivortex are close to
(far from) each other.
Fig.~\ref{fig_init-velocity-vs-xd} displays this competition by plotting the
initial velocity, $y'_1(0)=y'_2(0)$, against the initial dipole separation,
$x_{\rm d} = 2x_1(0)$.  The initial velocity is negative (positive) for
smaller (larger) $x_{\rm d}$.

We also note that the vortex dipole initial velocity approaches that of a
single vortex, placed at $x_1(0) =x_{\rm d}/2$, when the distance is large
enough, demonstrating that the motion of the two defects are indeed determined
primarily by the trap shape when they are far enough apart. 
%

At some intermediate distance, the two effects balance each other,
$y'_{1,2}(0)=0$.  This nontrivial \emph{stationary configuration} is a central
theme of this work.  We denote as $x_{\rm s}$ the stationary value of
$x_1(0)=x_{\rm d}/2$.  We explain later that above a critical interaction
$g=g_{\rm c}\approx18.0$, a vortex dipole configuration is a true stationary
solution of the full Gross-Pitaevskii solution such that not only the
singularity positions but also the density is stationary.

Fig.~\ref{fig_xs-vs-g} shows the stationary point positions $x_s$ as functions
of the interaction strength $g$, calculated using the trial wavefunctions
\eqref{trialwf_gaussian} and \eqref{trialwf_TF}.  Also shown are numerically
exact values of $x_s$, calculated directly from the GPE
(Sec.~\ref{sec_soliton}) for $g>g_{\rm c}\approx{18.0}$.  The comparison shows
that the gaussian ansatz is not useful in the relevant region $g>g_{\rm c}$.
The Thomas-Fermi ansatz \eqref{trialwf_TF} gives good results for the static
solutions until $g\sim{90}$.  We will therefore restrict ourselves to this
wavefunction in Sec.~\ref{sec_variational-trajectories}.

For very large $g$, the ansatz \eqref{trialwf_TF} yields decreasing behavior
for the $x_{\rm s}(g)$ function, presumably due to the `large-vortex' nature
of our wavefunction.

\subsection{Trajectories}  \label{sec_variational-trajectories}

\begin{figure}
\centering
\includegraphics*[width=0.9\columnwidth]{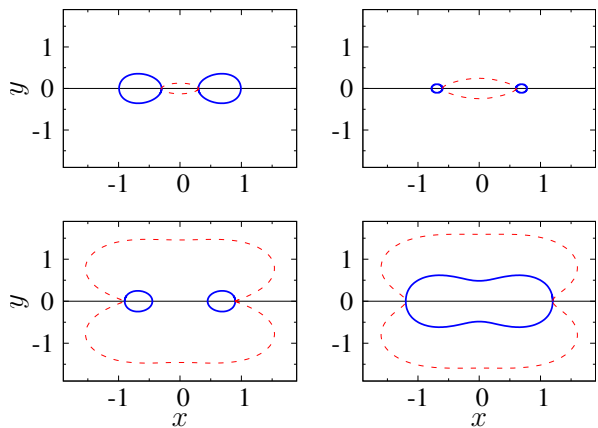}
\caption{ \label{fig_tf-trajectories}
(Color online.)  Vortex and antivortex trajectories, calculated using the
  variational formalism with wavefunction \eqref{trialwf_TF}.  The four
  pictures show trajectories resulting from initial positions $x_1(0)= $ 0.3,
  0.6, 0.9, 1.2. 
Full and dashed lines show $g=6$ and $g=40$ respectively.  
Lengths are plotted in trap units.
}
\end{figure}

The full curves of Fig.~\ref{fig_tf-trajectories} are vortex pair
trajectories, calculated
using the Thomas-Fermi wavefunction \eqref{trialwf_TF} for $g=6$.  
When placed close together ($x_1(0)<x_{\rm s}$), the pair starts by moving in
the negative $y$ direction, due to mutually-driven dipole motion.  As they
move away from the $x$-axis, inhomogeneity-driven motion becomes more and more
important, and eventually they turn away from each other and move in the
positive $y$ direction, crossing the $x$-axis at some $x_1 > x_{\rm s}$.  As
they move in the $+y$ direction, the distance between them decrease, and this
causes the direction to reverse again.  The reversals of motion might be
regarded as `reflections' of the propagating vortex dipole, at each `end' of
the trap.  
%

\begin{figure}
\centering
\includegraphics*[width=0.85\columnwidth]{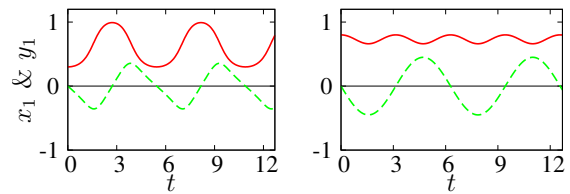}
\caption{  \label{fig_position-vs-time}
(Color online.)  Vortex position ($x_1$,$y_1$) as function of time.  Left
  panel: variational calculation with Thomas-Fermi ansatz ($g=6$).  Right
  panel: non-interacting bosons solved in terms of harmonic oscillator
  wavefunctions.     
All quantities are plotted in trap units.
}
\end{figure}

The variational calculation gives periodic orbits for vortex and antivortex
(first three panels of Fig.~\ref{fig_tf-trajectories}; left panel of
Fig.~\ref{fig_position-vs-time}).
When the starting positions $x_1(0)$ are moderately larger than $x_{\rm s}$, a
similar closed orbit pair results, starting from the outer edges of the closed
trajectories.  The closed orbits are smaller if the initial distance from the
stationary point, $|x_1(0)-x_{\rm s}|$, is smaller.
%
%
The dashed lines and the last panel in Fig.~\ref{fig_tf-trajectories} show
unphysical features in trajectories with large $g$ or large $x_1(0)$, which
are artifacts of our simple variational setup.

We will show in Sec.~\ref{sec_tdgpe} that the full Gross-Pitaevskii solution
supports very similar trajectories for large $g$.  
Our simple variational approach has thus already uncovered a very
characteristic motion of vortex dipoles in 2D trapped condensates, namely,
trajectories in which the two defects each revolve around a stationary point.


%
%



\section{Non-interacting bosons}  \label{sec_noninteracting}

The case $g=0$, \emph{i.e.}, non-interacting bosons in a harmonic trap, is
exactly solvable.  One can build any desired initial state as a linear
combination of appropriate harmonic oscillator eigenstates.  The evolution of
each eigenstate is known --- an eigenstate with energy $E$ evolves as
$\phi(t)=\phi(0)e^{-iEt}$.  Therefore the evolution of any state can be found
once it is expressed in this basis.

A state will be \emph{stationary} only if it is built out of eigenstates with
the same energy $E$.  Considering the wavefunctions of a 2D harmonic
oscillator (\emph{e.g.}, Ref.~\cite{book_cohen-tannoudji}), it is easy to
convince oneself that
an off-center vortex-antivortex pair cannot be formed out of the eigenstates
corresponding to a single energy.  Thus a stationary vortex dipole
configuration does not exist for $g=0$.

However, one can form a quasi-stationary state in which the defect positions
are stationary, although the condensate density undergoes complicated periodic
motions.  This is formed by combining eigenstates from the three lowest
eigenvalues, so that there are off-center vortex and antivortex at
($\pm{x}_1(0)$,$0$).  It is straightforward to show $y'_1(0)=
x_1(0)-1/x_1(0)$, also plotted in Fig.~\ref{fig_init-velocity-vs-xd}.  The
quasi-stationary configuration is therefore $x_{\rm s}=1$.

The defect trajectories for $g=0$, found from the above linear combination,
involve sharp direction reversals along the same path.  This suggests that the
vortex picture is not very suitable here.
In fact, the density profiles do not show
well-defined vortex and antivortex, but a `soliton'-like band of low density,
even when the system contains a dipole-like pair of singularities at
($\pm{x}_1(0)$,$0$).

The right panel of Fig.~\ref{fig_position-vs-time} shows the defect
trajectories for the 2D non-interacting trapped condensate, via the functions
$x_1(t)$ and $y_1(t)$.  The $x_1(t)$ curve has half the period of $y_1(t)$,
indicating that the defects move back and forth along the same line instead of
around a closed orbit.

\section{Static solutions: solitons \& vortex dipoles}  
\label{sec_soliton}

\begin{figure}
\centering
\includegraphics*[width=0.95\columnwidth]{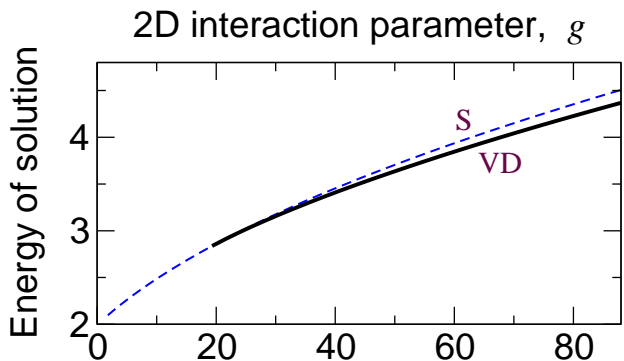}
\caption{ \label{fig_bifurcation-energies}
(Color online.) The stationary soliton solution bifurcates at $g{\approx}18$,
  producing a stationary vortex dipole solution.  Dashed (full) line shows
  energy of a stationary soliton (stationary vortex dipole).}
\end{figure}

The absence of a vortex-antivortex stationary solution for $g=0$, and some of
the dynamics to be presented in Sec.~\ref{sec_tdgpe}, can be put into context
by considering a bifurcation phenomenon \cite{crasovan-etal_PRA03}.
%

At small $g$, we do have a stationary solution: a dark soliton-like object
instead of a vortex dipole.  This is easy to see for $g=0$: the first-excited
($n_x$,$n_y$) =(0,1) eigenstate, $\psi_{01}\propto ye^{-(x^2+y^2)/2}$, can be
regarded as a dark soliton analog.
The term `soliton' comes from the analogy to the case where the confinement in
the $y$ direction is much weaker, \emph{i.e}, a `quasi-1D' condensate.
Although the word evokes a picture of a propagating object, in a trapped
condensate a soliton can be stationary.

As $g$ is increased, the soliton-like configuration remains a stable
stationary object until $g=g_{\rm c}$, at which point it bifurcates into two
solutions, of which the unstable one is still soliton-like and the stable one
is the vortex dipole configuration.  Using a many-variable Newton-Ralphson
method to find solutions of the time-\emph{independent} GPE, we have found
that the point above which the stationary vortex dipole exists is $g_{\rm
c}\approx 18$.  This is consistent with Ref.~\cite{crasovan-etal_PRA03}, but
provides better precision.  The energies of the soliton solution and the
vortex dipole solution are shown in Fig.~\ref{fig_bifurcation-energies} as a
function of $g$, and density profiles for stable stationary solutions are in
Fig.~\ref{fig_stationary-density-plots}.

The same bifurcation happens in elongated traps, where the word `soliton' is
more appropriate.  In a 3D trap one gets via bifurcation a stationary
\emph{vortex ring} instead of a vortex dipole
\cite{KomineasPapanicolaou_LaserPhys04}.

The filled circles in Fig.~\ref{fig_xs-vs-g} indicate stationary vortex dipole
positions for $g>g_{\rm c}$.  Note that the value of $x_{\rm s}$ is near the
trap oscillator length ($x_{\rm s}\approx1$) over a wide range of $g$.  This
is well below the Thomas-Fermi size of the system (dotted line in
Fig.~\ref{fig_xs-vs-g}), which increases as $R_{\rm TF}\sim{g}^{1/4}$ with the
interaction.

\section{Numerical evolution of Gross-Pitaevskii equation}  \label{sec_tdgpe}

\begin{figure}
\centering
\includegraphics*[width=0.3\columnwidth]{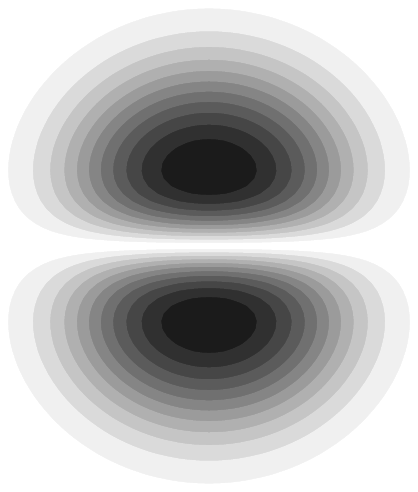}
\includegraphics*[width=0.3\columnwidth]{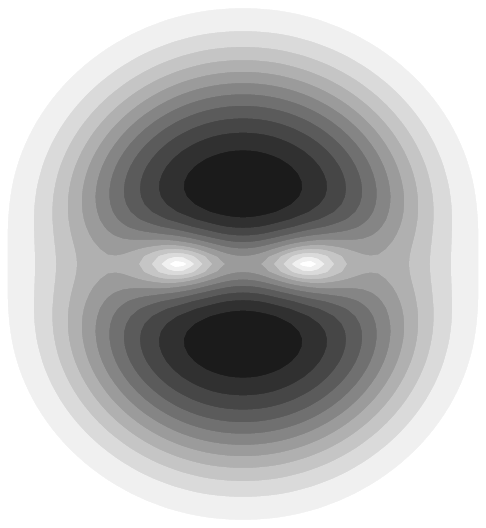}
\includegraphics*[width=0.3\columnwidth]{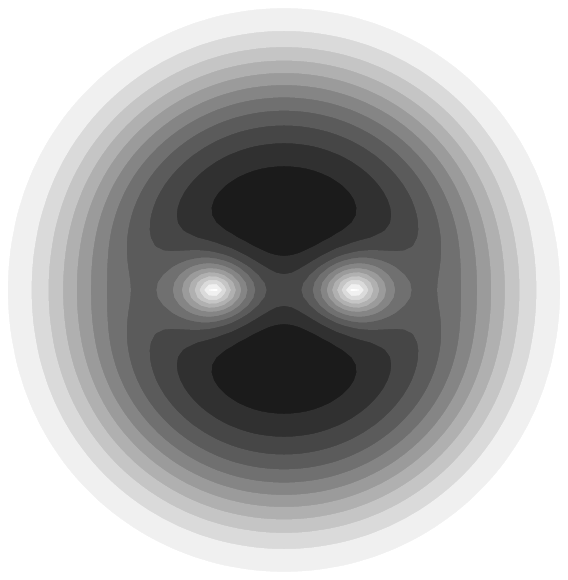}
\caption{ \label{fig_stationary-density-plots}
Density profiles of stationary states: (1) soliton at $g=11$; (2) vortex
dipole at $g=25$; (3) vortex dipole at $g=60$.  
}
\end{figure}

%
We now present results from explicit numerical evolutions of the 2D
time-dependent Gross-Pitaevskii equation.  In these simulations, a vortex
dipole was first created at $x=\pm{x}_{\rm d}/2$ by propagating in imaginary
time, at each step normalizing and also imposing the phase structure of
$(z-z_1)(z^*-z^*_2)$.  After reaching a stable dipole configuration,
propagation in real time was followed.  Time propagation was performed using
the 4th-order Runge-Kutta algorithm, with the kinetic energy term calculated
in momentum space.

Since we have no dissipation, the energy should remain constant.  Breakdown of
the numerical scheme, which always happens after some amount of propagation
time due to finite spatial and temporal grids, is signaled by a rapid increase
of the energy.  The simulation breaks down earlier for larger values of
interaction $g$, because the length scale $\xi\sim{g}^{-1/4}$ is smaller.
With spatial grids of up to $\sim{100}$ per trap oscillator length, we
obtained useful results up to $g\lesssim{200}$.

Locating the vortex and antivortex is tricky, especially for smaller $g$.  It
is then often not sufficient to analyze the densities. One possibility is to
observe both real and imaginary parts of $\psi$ and find points where they
both vanish; another tactic is to track phases and identify plaquettes around
which there is a $2\pi$ phase winding.  

Our main observation is that the motion of the vortex dipole is never
periodic, as opposed to the variational description.  For small $g$, the
motion is far more complicated.  For larger $g$, the vortex pair starts to
make almost closed orbits.  Fig.~\ref{fig_stationary-density-plots} shows one
of these relatively easier cases, at $g=150$, where we have followed the
dynamics for slightly longer than one `period'.  As in the variational
calculation, the competition between mutually driven and inhomogeneity-driven
motion is clear, as is the tendency to reverse direction when the vortex
dipole reaches the `edge' of the condensate.  There are several additional
unexplained features.  One is the structure at the outer edges of the orbits.
Also, for large initial pair distances $x_1(0)>x_{\rm s}$, we note that the
defects first move in the $-y$ direction for a very short time, before
starting in the $+y$ direction.  These two features might be due to effects of
the relatively nearby boundary \cite{Anglin_PRA02, AlKhawaja_PRA05}.

The situation is significantly more complicated for smaller $g$, as shown in
Fig.~\ref{fig_gp-orbits_smaller-g}.  The vortex positions are joined by
straight lines.  The big jumps in these `trajectories' represent intervals
where the singularities were difficult to track or distinguish, or there were
additional singularity pairs.  The vortex and antivortex in the $g=50$ case
each make almost one complete revolution around the stationary point, before
coming too close to each other, and after this the situation for some time
becomes too complicated to track.  This is when the pair undergoes some type
of `annihilation', leading to some complicated excitations including
additional singularity pairs.  When we next manage to track a pair, they move
again roughly along trajectories around the stationary point, until they come
near each other again.

The $g=10$ case is even more complicated.  For $g<18$, there is no true
stationary vortex dipole solution; therefore the paradigm of each defect
rotating around a stationary position is not expected to provide a useful
description.  Nevertheless, some remnant of the pair `revolving around
stationary positions' seems to be visible in the defect trajectories.

\begin{figure}
\centering
\includegraphics*[width=0.93\columnwidth]{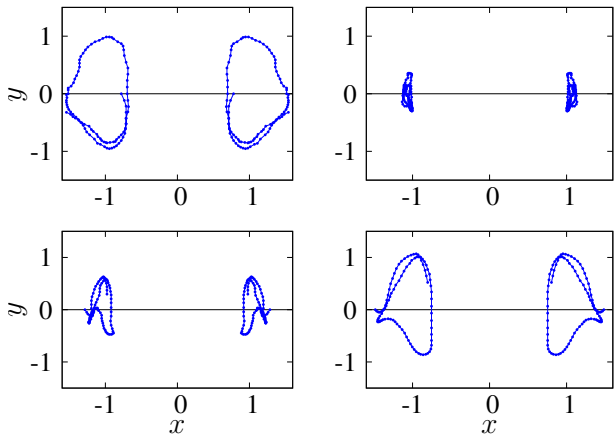}
\caption{ \label{fig_gp-orbits_g150}
(Color online.) Vortex pair trajectories from numerical solution of
  time-dependent GPE, $g=150$.  Starting positions are $x_1(0)=$ 0.8, 1.1,
  1.3, and 1.5.  Successive vortex/antivortex positions have been joined by
  straight lines.  The condensate radius (Thomas-Fermi) is $R_{\rm TF}\simeq 3.72$,
  not visible in these plots. 
Lengths are plotted in trap units.
}
\end{figure}

\section{Densities}  \label{sec_densities}

\begin{figure}
\centering
\includegraphics*[width=0.95\columnwidth]{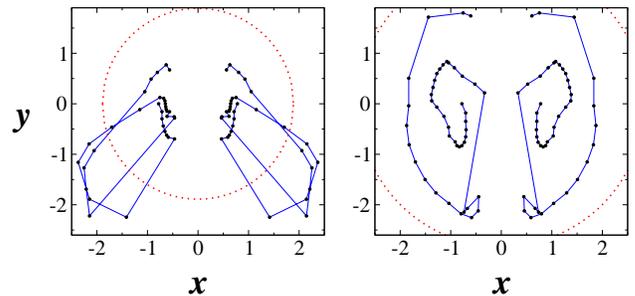}
\caption{ \label{fig_gp-orbits_smaller-g}
(Color online.) Vortex pair trajectories from numerical evolution of
time-dependent GPE, $g=10$ and $g=50$.  Starting positions are $x_{1,2}(0)=
{\pm}0.8$.  
Successive positions have been joined by straight lines.  There are some gaps
in time (widely separated points joined by a straight line).  These correspond
to intervals where the presence of several vortex-antivortex pairs made it
difficult to meaningfully track a single pair.
The Thomas-Fermi value for the condensate radius are $R_{\rm TF}\simeq 1.89$
and $R_{\rm TF}= \simeq 2.825$ in the two cases, shown with dotted lines.
Lengths are plotted in trap units.
}
\end{figure}

It is only at rather large $g$, when the vortex size $\xi$ is small, that
density plots are useful for identifying vortices.  The middle figure in
Fig.~\ref{fig_stationary-density-plots} shows that for $g$ moderately above
$g_{\rm c}\approx18$, the density profiles for stationary vortex dipole
configurations look substantially like the stationary soliton case
(Fig.~~\ref{fig_stationary-density-plots} left).  Obviously in a dynamic
situation at such interactions, the vortices retain little of their individual
identity as they move around, which helps explain why the singularity
trajectories in Fig.~\ref{fig_gp-orbits_smaller-g} are not easy to explain in
the vortex-antivortex language.

For $g{\sim}$50-100, the vortices are more well-defined in stationary density
profiles (Fig.~\ref{fig_stationary-density-plots} right), but still have
significant shape dynamics when they are moving.  It is only at very large $g$
that the vortex and antivortex can be regarded as point objects.  This is the
case where the Lagrangian analysis, based on wavefunctions with well-defined
vortices, gives good insight.

Density plots of the non-interacting Bose condensate with a `vortex dipole'
(Sec.~\ref{sec_noninteracting}) typically show a `soliton'-like profile rather
than distinct vortex and antivortex, indicating that the vortex picture is of
limited relevance.

Our variational wavefunctions \eqref{trialwf_gaussian} and \eqref{trialwf_TF}
also have large vortex sizes.  As a result, when the pair distance is not very
large, the density profile tends to show a large stripe of low-density region
(\emph{i.e.}, soliton-like profile) instead of individual depressions for
vortex and antivortex.  However, the distinct identities of the vortex and
antivortex (as well as lack of additional dynamics possibilities) is built
into these wavefunctions.

\section{Discussion}  \label{sec_discussion}

To summarize, we have analyzed, using several methods and viewpoints, the
trajectories of a vortex-antivortex pair in a circularly trapped Bose-Einstein
condensate, described by the Gross-Pitaevskii equation.  The principal
intuition we have provided is in terms of the competition between
mutually-driven and inhomogeneity-driven motions of the pair.  This leads to a
simple understanding of the stationary vortex dipole solutions
\cite{crasovan-etal_PRA03, Helsinki_stationary-clusters_PRA05}, and also leads
to the expectation of a type of motion where the vortex and antivortex each
revolve around the stationary positions.
Although our time-evolution simulations do not extend to interactions beyond
$g=200$, our available results strongly indicate that this type of pair motion
becomes more and more relevant for larger interactions.

The fact that low-$g$ condensates show more complicated vortex dynamics has
been given proper context through our considerations of the non-interacting
case, the stationary solutions below and above $g=g_{\rm c}\approx{18}$, and
density profiles, in Secs.\ \ref{sec_noninteracting}, \ref{sec_soliton},
\ref{sec_densities}.  
At large $g$, the vortex shape distortion dynamics is less important (because
they are smaller objects), there is lesser tendency of pairs to morph into or
mimic soliton-like objects, extra pairs are not created as easily, and
annihilation is less likely in the absence of dissipation.  All this favors
the type of motion described above, obtained from our variational wavefunction
which only allows position dynamics for a single vortex dipole.  At smaller
$g$, the description in terms of vortex positions alone is less adequate,
because of additional dynamics possibilities.

Vortex dipoles can be regarded as the two-dimensional analogs of dark solitons
in one dimension, and of vortex rings in three dimensions.  Each of these are
solitary waves in uniform (untrapped) Bose condensates of the appropriate
dimension, and can become non-propagating stationary objects in trapped
condensates.  Vortex rings in trapped 3D BECs display self-propelled and
inhomogeneity-driven motions in opposite directions, just as we have found for
vortex dipoles in 2D.

Our study opens up a number of additional questions.  \\ 
1.  As described in Sec.~\ref{sec_tdgpe}, there are aspects of the
trajectories that we do not understand even for relatively large $g$, such as
the initial negative-$y$ motion for $x_1(0)>x_{\rm s}$
(Fig.~\ref{fig_gp-orbits_g150}).  Presumably, this requires understanding of
the coupling of the vortex position dynamics to the vortex shape dynamics.  It
is intriguing to ask whether the trajectory becomes smoother and more periodic
at even larger $g$.  \\
2. For small $g$, an additional open question is how the vortex dipole motion
is affected by the presence of `nearby' solitonic solutions, \emph{e.g.}, this
might explain the appearance of extra pairs.  (A soliton may be regarded as a
very large number of vortex dipoles.)  \\
3.  The `reflection' of the dipole when it reaches the `edge' of the condensate
could be better studied in the setting of an elongated 2D condensate, where
edges are better defined than in our circular case.  \\
4. There remains the open question of vortex dipoles which are initially not
placed symmetrically around the trap center. \\
5. We have not studied the effects of dissipation on vortex dipole dynamics.
In particular, dynamics of vortex pair creation and annihilation is important
for Kosterlitz-Thouless physics \cite{Dalibard_KT-expts}.  At present we do
not have a satisfactory theoretical framework for including such effects.  

We hope to address some of these issues in future work.

\appendix*

\section{Variational formalism}

We describe here the variational Lagrangian formulation used in
Sec.~\ref{sec_variational} to follow the motion of vortex and antivortex
positions.
%
%
Our results are based on variational wavefunctions of the form
\eqref{trialwf_general}, where vortex positions appear as time-dependent
variational parameters $z_1(t)$ and $z_2(t)$.
Using such a wavefunction in the Lagrangian
\begin{multline}   \label{lagrangian}
L=\int dr\bigg[\ \frac{i}{2} \left(
\psi^*\frac{\partial\psi}{\partial{t}} -\psi\frac{\partial\psi^*}{\partial{t}}  
\right) 
\\
+~ \tfrac{1}{2}\psi^*\bigtriangledown^2\psi 
  ~- V_{\rm tr}(\vr)\ |\psi|^2 ~- \hf g\ |\psi|^4 \bigg]  \; ,
\end{multline}
one can derive Euler-Lagrange equations of motion for the complex parameters
$z_1(t)$ and $z_2(t)$.  Solving the equations of motion then provides the
trajectories of the topological defects.  This method for vortex dynamics in
BECs was pioneered in Ref.~\cite{Fetter_PRL84} and has since been used in
several vortex applications \cite{LundhAo_PRA00, LinnFetter_PRA00,
Fetter_annular_2papers, Bhattacherjee_JPB04, AlKhawaja_PRA05, SnoekStoof}.


For the condensate shape function $f_{\rm c}$ of Eq.~\eqref{trialwf_general},
we used both Gaussian and Thomas-Fermi forms:
\begin{equation}
\label{trialwf_gaussian}
\psi_{\rm G} ~=~ A_{\rm G}(t) \; \lbc{z-z_1(t)}\rbc\lbc{z^*-z^*_2(t)}\rbc \; \exp\lbc-|z|^2/2\rbc
\end{equation}
and
\begin{equation}
\label{trialwf_TF}
\psi_{\rm TF} ~=~ A_{\rm TF}(t) \; \lbc{z-z_1(t)}\rbc\lbc{z^*-z^*_2(t)}\rbc \;
f_{\rm TF}
\;,
\end{equation}
with $f_{\rm TF}= \sqrt{\lba\mu-\hf|z|^2\rba/g}$ and $\mu=\sqrt{g/\pi}$.  
%
%

Here $A(t)$ is a time-dependent normalization constant.  Retaining this factor
$A(t)$ is essential for getting the dynamics even qualitatively correct.  For
example, for the single-vortex case, omitting this factor gives vortex
precession in the \emph{reverse} direction!  Since $A(t)$ actually contains
the vortex position parameters $z_{1,2}(t)$, terms in the equations of motion
are missed when $A(t)$ is omitted.

Our form \eqref{trialwf_general} is generalized from the lowest-Landau-level
form for vortices in a rapidly rotating condensate \cite{tlho-prl87}.  It has
the advantage of being amenable to exact treatment, \emph{i.e.}, the
integrations in \eqref{lagrangian} can be performed analytically.  The
disadvantage is that the vortices are `too big', as discussed in
Sec.~\ref{sec_variational}.
We could in principle use any wavefunction of the form
\begin{equation}
\psi ~=~ A(t) \; g_{\rm v}(u_1) e^{i\phi_1} \; g_{\rm v}(u_2) e^{-i\phi_2} \;
f_{\rm c}(|z|^2)
\end{equation}
where $u_i = |z-z_i|/\xi$ and $\phi_i = \tan^{-1}(\frac{y-y_i}{x-x_i})$.
The vortex shape function $g_{\rm v}(u)$ ideally would be linear in $u$ for
$u\ra{0}$ and constant for large $u$ \cite{Fetter_PRA65-I}; for example
\[
g_{\rm v}(u) =\frac{u}{\sqrt{u^2+\xi^2}} 
\qquad {\rm or} \qquad 
g_{\rm v}(u) =1-e^{-u/\xi} \, .
\]
Eq.~\eqref{trialwf_general} corresponds to $g_{\rm v}(u)=u$, having the wrong
behavior at large $u$.  This is why our variational wavefunctions have
unnaturally large vortices.  Unfortunately, we were unable to identify a
function $g_{\rm v}(u)$, having the correct limiting behaviors, for which the
integrals in Eq.~\eqref{lagrangian} can be done analytically.  We also did not
attempt to incorporate velocity field effects due to the boundary
\cite{Anglin_PRA02, AlKhawaja_PRA05}.
Our presented results are therefore limited to wavefunctions of the form of
Eq.~\eqref{trialwf_general}.


Using one of our variational wavefunctions, \eqref{trialwf_gaussian} or
\eqref{trialwf_TF}, the Euler-Lagrange equations of motion, $D_t
\lba\partial{L}/\partial{\dot{u}}\rba = \partial{L}/\partial{u}$, become
equations of motion for the defect positions $x_{1,2}(t)$ and $y_{1,2}(t)$.
We can obtain these equations of motion analytically, but they are too
cumbersome to reproduce here.  

For a single vortex-antivortex pair with unrestricted positions, there are
four variables, leading to four coupled first-order nonlinear differential
equations which can be solved numerically for the vortex trajectory.  For the
initial conditions we have used, we expect the motion to be symmetric around
the $y$-axis, so that we can use $x_1(t)=-x_2(t)$ and $y_1(t)=y_2(t)$.  This
significantly reduces the numerical cost of solving the equations of motion,
of which there are now only two.


\acknowledgments

MH thanks Henk Stoof for discussions and insights.

\end{document}